\documentclass[
]{article}
\usepackage{lmodern}
\usepackage{xcolor}
\usepackage[margin=1.0in]{geometry}
\usepackage{amssymb,amsmath}
\usepackage{float}
\usepackage{ifxetex,ifluatex}
\usepackage[htt]{hyphenat}
\usepackage[colorlinks=true,
            linkcolor = blue,
            urlcolor  = blue,
            citecolor = blue,
            anchorcolor = blue]{hyperref}

\newcommand{\linkto}[2]{{#2}\footnote{\url{#1}}}

\ifnum 0\ifxetex 1\fi\ifluatex 1\fi=0 
  \usepackage[T1]{fontenc}
  \usepackage[utf8]{inputenc}
  \usepackage{textcomp} 
\else 
  \usepackage{unicode-math}
  \defaultfontfeatures{Scale=MatchLowercase}
  \defaultfontfeatures[\rmfamily]{Ligatures=TeX,Scale=1}
\fi
\IfFileExists{upquote.sty}{\usepackage{upquote}}{}
\IfFileExists{microtype.sty}{
  \usepackage[]{microtype}
  \UseMicrotypeSet[protrusion]{basicmath} 
}{}
\makeatletter
\@ifundefined{KOMAClassName}{
  \IfFileExists{parskip.sty}{%
    \usepackage{parskip}
  }{
    \setlength{\parindent}{0pt}
    \setlength{\parskip}{6pt plus 2pt minus 1pt}}
}{
  \KOMAoptions{parskip=half}}
\makeatother
\usepackage{graphicx}
\usepackage{authblk}
\makeatletter
\def\maxwidth{\ifdim\Gin@nat@width>\linewidth\linewidth\else\Gin@nat@width\fi}
\def\maxheight{\ifdim\Gin@nat@height>\textheight\textheight\else\Gin@nat@height\fi}
\makeatother
\setkeys{Gin}{width=\maxwidth,height=\maxheight,keepaspectratio}
\makeatletter
\def\fps@figure{htbp}
\makeatother
\providecommand{\tightlist}{%
  \setlength{\itemsep}{0pt}\setlength{\parskip}{0pt}}
\DeclareFontFamily{\encodingdefault}{\ttdefault}{\hyphenchar\font=`\-}
\DeclareTextFontCommand{\mytexttt}{\ttfamily\hyphenchar\font=45\relax}

\begin{document}

\title{\textbf{} Unity is Strength: A Formalization of Cross-Domain Maximal Extractable Value}
\author[1]{Alexandre Obadia}
\author[2]{Alejo Salles}
\author[3]{Lakshman Sankar}
\author[4]{Tarun Chitra}
\author[5]{Vaibhav Chellani}
\author[6]{Philip Daian}

\affil[1,2,6]{Flashbots Research (\textit{\{alex, alejo, phil\}@flashbots.net})}
\affil[3]{Ethereum Foundation (\textit{lsankar4033@gmail.com})}
\affil[3]{Gauntlet (\textit{tarun@gauntlet.network})}
\affil[5]{Movr Network (\textit{vaibhav@movr.network})}

\maketitle

\begin{abstract}
The multi-chain future is upon us. Modular
architectures are coming to maturity across the ecosystem to scale
bandwidth and throughput of cryptocurrency. One example of such is the
Ethereum modular architecture, with its beacon chain, its execution
chain, its Layer 2s, and soon its shards. These can all be thought as
separate blockchains, heavily inter-connected with one another, and
together forming an ecosystem.

In this work, we call each of these interconnected blockchains
`domains', and study the manifestation of Maximal Extractable Value
(MEV, a generalization of ``Miner Extractable Value'') across them. In
other words, we investigate whether there exists extractable value that
depends on the ordering of transactions in two or more domains jointly.

We first recall the definitions of Extractable and Maximal Extractable
Value, before introducing a definition of Cross-Domain Maximal
Extractable Value. We find that Cross-Domain MEV can be used to measure
the incentive for transaction sequencers in different domains to collude
with one another, and study the scenarios in which there exists such an
incentive. We end the work with a list of negative externalities that
might arise from cross-domain MEV extraction and lay out several open
questions.

We note that the formalism in this work is a work-in-progress, and we
hope that it can serve as the basis for formal analysis tools in the
style of those presented in Clockwork Finance~\cite{babel2021clockwork}, as well as for discussion
on how to mitigate the upcoming negative externalities of substantial
cross-domain MEV.

\end{abstract}

\hypertarget{introduction}{%
\subsection{1 Introduction}\label{introduction}}

MEV was originally defined in~\cite{daian2020flash} to study the effects of application-layer activity on consensus-level incentive perturbations, as well as to inform better application design. To understand how this concept of MEV may affect a cross-domain, multi-blockchain future, including how the incentives of this future may be threatened or destabilized, we must first
define the concept of a \emph{domain}.

\hypertarget{what-are-domains}{%
\subsubsection{1.1 What are domains?}\label{what-are-domains}}

\textbf{definition.} A \emph{domain} is a self-contained system
with a globally shared state.  This state is mutated by various players through
actions (often referred to as ``transactions''), that execute in a shared execution environment's semantics.

\textbf{definition.} A \emph{sequencer} is an actor that can
control the order of actions within a domain before they are executed,
and thus influence future states of this domain.

Layer 1s, Layer 2s, side-chains, shards, centralized exchanges are all
examples of domains. For the purpose of this work, there is no need to
understand how these systems work, as we will abstract such details away.

Often, in a blockchain, transactions are applied sequentially to a domain's current state, according to the state transition function of the domain. Rather than transactions, we use the concept of actions in the rest of this work to represent their impact on the state to avoid confusion and force generality, as some domains, such as centralized exchanges, may experience state changes that are not effected by ACID-style transactions (Atomicity, Consistency, Isolation, and Durability).

\textbf{definition.} An \emph{action} \(a\) is a mapping of a state to
another state.

We write the effect of a sequence of actions \(a_1 \ldots a_n\) on
a state \(s\) as:

\[{\atop s\; \xrightarrow[{a_1 \dots a_n}] \;s'} \]

where \(s'\) is the state arrived at after the sequence of actions has
been applied to \(s\).

\hypertarget{motivation-multi-domain-arbitrage}{%
\subsubsection{1.2 Motivation: multi-domain
arbitrage}\label{motivation-multi-domain-arbitrage}}

We will now motivate our definition by providing an example of cross-domain Extractable Value that exists in the wild.  We will then extend the definition in~\cite{babel2021clockwork} to capture the changes in economic incentives introduced by this cross-domain MEV, and use this example to illustrate the power and generality of our definition.

Applications such as automated market makers and lending markets are
being instantiated on each new domain that sees the light of day.
\begin{figure}[H]
    \centering
    \includegraphics[width=0.8\textwidth]{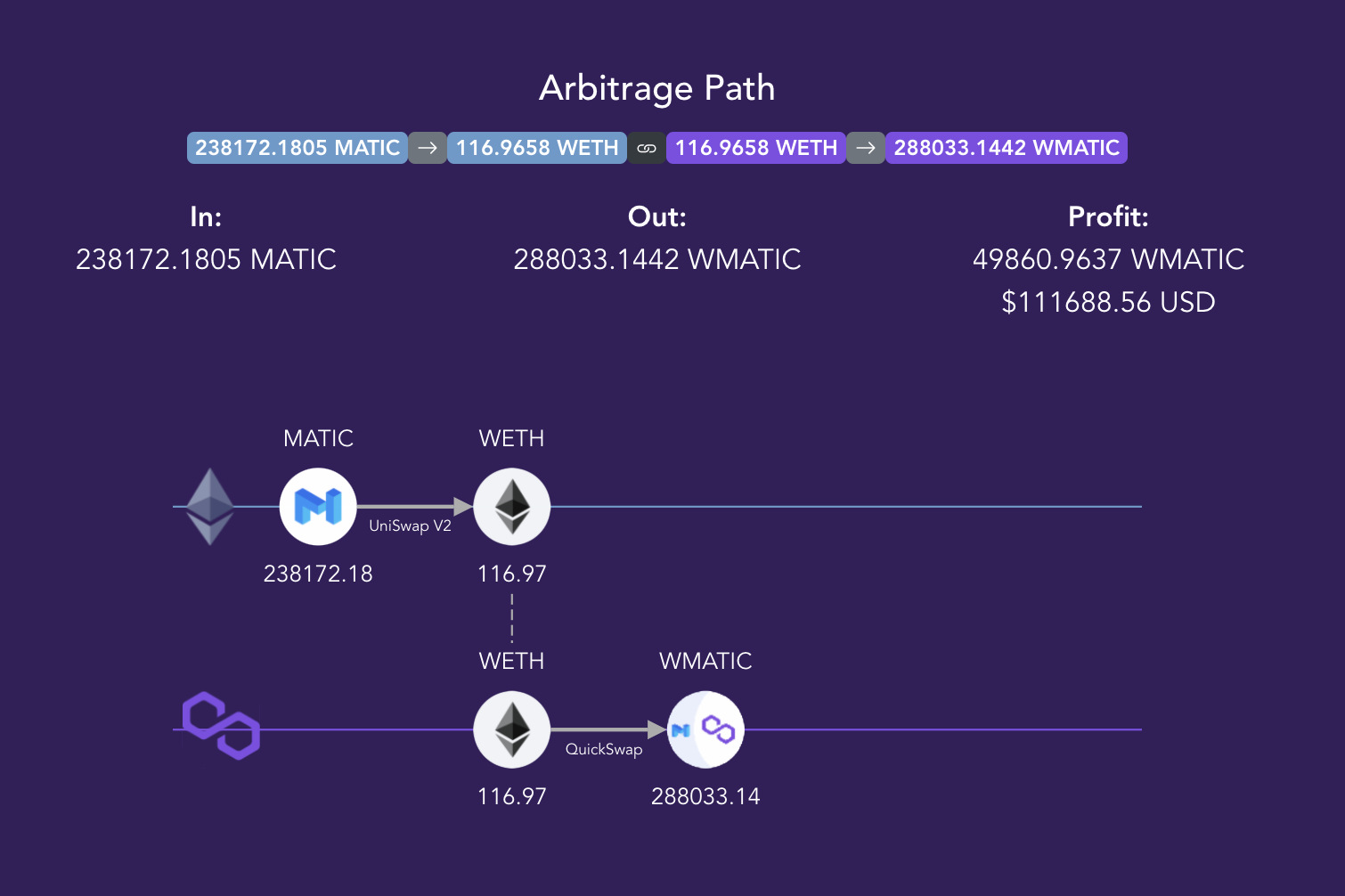}
    \caption[]{Example of 2-domain arbitrage between Ethereum and Polygon.  Source: \linkto{https://westerngate.xyz/}{Westerngate}}
    \label{fig:2domain}
\end{figure}
\footnotetext{https://westerngate.xyz/}

For example, the popular automated market maker Sushiswap with about
\$3B weekly volume on Ethereum \footnote{\url{https://dune.xyz/nascent/SushiSwap}}, started by existing only in
Ethereum
\linkto{https://github.com/sushiswap/sushiswap/commit/842679342047ecd0e0126f05b6089f8b727cb063}{a
year ago}, and now exists also on
\linkto{https://twitter.com/josephdelong/status/1367226781393166336?s=20}{XDai,
Polygon, Fantom},
\linkto{https://twitter.com/SushiSwap/status/1432987097783103496?s=20}{BSC,
Arbitrum, Avalanche, Celo, Palm and Harmony}.

This means there likely exist liquidity pools representing the same asset pairs
on each domain, yet with different volume, depth and activity. So, there
will be a point in time where pools in different domains for the same
asset pairs will be relatively imbalanced, creating an arbitrage
opportunity.

Figure~\ref{fig:2domain} shows an example of one such opportunity observed by a real-time arbitrage detection tool. In this instance, Uniswap V2 on Ethereum was offering the ability to swap 238172.18 Matic for 116.97 WETH, which could then be swapped for 288033.14 WMATIC on Polygon. Any trader which values WMATIC and MATIC at the same 1:1 price ratio (a reasonable assumption, as MATIC and WMATIC can be exchanged at 1:1 over a bridge with a small time penalty and fee) would have profited 49860.9637 MATIC for bridging this price gap, a net profit of \$111,688 USD. Such opportunities will occur any time a user trades on \emph{any AMM} without perfectly splitting their trade across all pools, a direct extension of the 2-AMM instability result in~\cite{babel2021clockwork}.

\begin{figure}[H]
    \centering
    \includegraphics[width=0.8\textwidth]{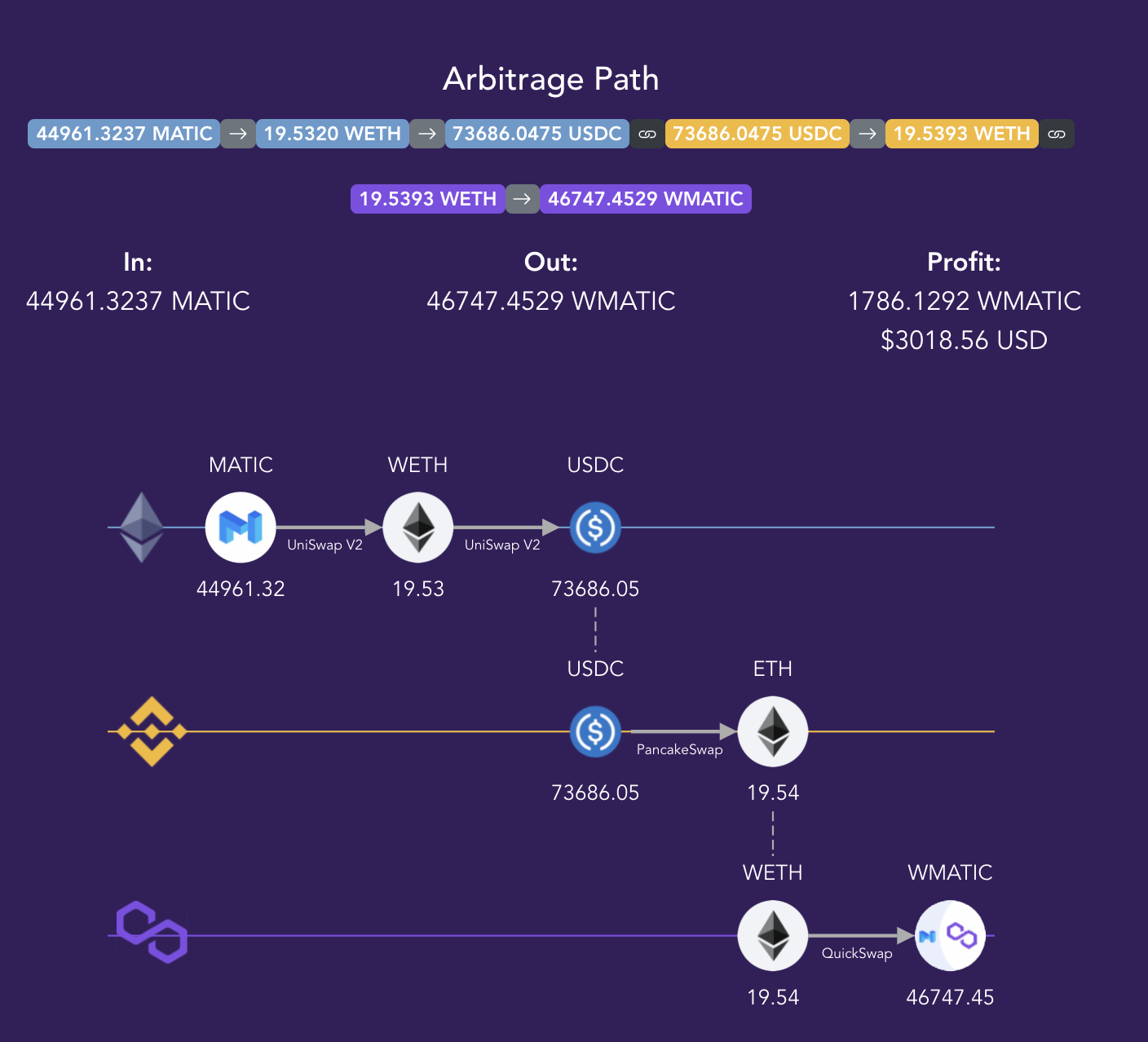}
    \caption{Example of 3-domain arbitrage between Ethereum, Binance Smart Chain, and Polygon. Source: \protect \linkto{https://westerngate.xyz/}{Westerngate}}
    \label{fig:3domain}
\end{figure}
\footnotetext{https://westerngate.xyz/}

This is not limited to the 2-domain case: Figure~\ref{fig:3domain} shows a similar opportunity identified between all three domains, resulting in a profit of \$3018.56. Unlike the 2-domain case, this extraction requires moving or bridging assets across three domains rather than two.

At first glance, these cross-domain opportunities look very similar to a pure-profit arbitrage opportunity that forms the basis for the classical notion of MEV in~\cite{daian2020flash}.
However, unlike in~\cite{daian2020flash}, the opportunity now depends on the ordering of transactions in
multiple domains rather than a single one, the assets our trader starts
with and ends with are different (Matic and WMatic), and the mechanics
by which this opportunity is seized seems to involve transferring assets
across domains, a non-atomic operation that breaks the atomicity described in~\cite{daian2020flash}.

There are many substantive differences to untangle, each with their own implications on consensus protocol design, user fairness, and the economics of all blockchain protocols and their MEV ecosystems. We will now attempt to capture these differences in a formal definition.

\hypertarget{defining-cross-domain-mev}{%
\subsection{2. Defining Cross-Domain
MEV}\label{defining-cross-domain-mev}}

Before introducing a definition of Cross-Domain Maximal Extractable
Value, we first introduce the concepts of \emph{reachable states},
\emph{single-domain extractable value} and \emph{single-domain maximal
extractable value}. We re-use definitions proposed in~\cite{babel2021clockwork}.

\hypertarget{reachable-states}{%
\subsubsection{2.1 Reachable States}\label{reachable-states}}

Each domain has its own state transition function with specific validity
rules. An example of a validity rule is that an account cannot spend
more than it has. The set of all valid actions a player $P$ can take represents
$P$'s power in the MEV extraction game on that domain.

We define a set of actions $A_P$ which represent all valid actions available to
player $P$.
We say a sequence of actions \(a_1...a_n\) $\in A_P$ when $\forall i$, $a_i \in A$ (where $A$ is the global set of all actions for all players in the protocol), and $\forall i, j$, $a_i \neq a_j$ (each action is allowed and unique). For example, pending transactions in the mempool able 
to be executed by a sequencer as described in~\cite{babel2021clockwork} may each form an available action in L1 transaction sequencing.

For each sequence of allowed actions, there exists a state \(s'\) that
will be the state of our domain in the future after each action is
applied sequentially to the current state $s$.

All \(s'\)s together are the \emph{reachable states} of our domain in
the future under the influence of player $P$ ($S'_P$), given its current state \(s\) and a set of actions. We define:

\[{s \xrightarrow[A] \ S_P'} \]

such that $S_P' = \{s' | s \xrightarrow[{a_1, \dots, a_n} \in A_P] \ s'\}$

In practice, the available set of actions $A$ may depend on the initial world state $s$ (for example, a validator may be able to include certain transactions only when $s$ has enough balance to pay their fees). We will thus consider $A$ to be defined as a function $A_P(s)$, but will omit this extra parameter for convenience in notation.

$S'_P$ captures which future states are reachable given the behavior of a specific player $P$. In the remainder of this work, we will often omit $P$ wherever it occurs in the notation for simplicity. In this case, we will assume $P$ is a generic player with an arbitrary address, and must define the capital available to this address in $s$ as well as the action space available to $P$ in a specific domain (e.g. re-ordering transaction as in~\cite{babel2021clockwork}).

\hypertarget{single-domain-extractable-value}{%
\subsubsection{2.2 Single-Domain Extractable
Value}\label{single-domain-extractable-value}}

\textbf{definition.} \emph{Extractable Value} is the value between one
or more blocks accessible to any user in a domain, given any arbitrary
re-ordering, insertion and censorship of pending or existing
transactions.

More formally, and using the terms we introduced above, we define a user
\(P\)'s \emph{extractable value} (\(ev_i\)) in domain \(i\) after executing a sequence of actions $a_1 ... a_n$ on an initial state $s$ as follows:

\[{ev_i(P,s,a_1 ... a_n)= b_i(s', P)-b_i(s, P) }\]

$\text{ where } {s \xrightarrow[a_1 ... a_n] \;s'} \text{and} \;{b_i(s,P)} \text{ is the balance of a player $P$ at a state s.}$

Note that we define the extractable value for a sequence of actions as the change in a player's balance across the state change between $s$ and $s'$, where this state change results from executing this sequence of actions. Our subscript here indicates in which domain a player's balance change is effected (that is, in which domain value is being extracted). This balance will likely be a numerical entry stored in a state, making the balance function for most domains a simple state lookup.

We note that in the original MEV definition in~\cite{babel2021clockwork}, miner-extractable value is provided as an upper-bound, and therefore considers actors P that are miners in Proof-of-Work Ethereum, the most privileged actors in the system. However, we transition below and throughout this paper to the notion of \emph{maximal} rather than miner-extractable value, to generalize to non-PoW chains and non-blockchain domains. Rather than considering specifically the action space of a special sequencer (or miner) that can reorder, insert, or censor, as in~\cite{babel2021clockwork}, we generalize our definition to allow custom action spaces for each player.

\hypertarget{single-domain-maximal-extractable-value}{%
\subsubsection{2.3 Single-Domain Maximal Extractable
Value}\label{single-domain-maximal-extractable-value}}

\textbf{definition.} \emph{Maximal Extractable Value} (MEV) is the
maximal value extractable between one or more blocks, given any
arbitrary re-ordering, insertion or censorship of pending or existing
transactions.

We can then simply take the maximum of our definition above over all valid sequences of actions for player $P$, essentially
looking for \emph{the} reachable state where our balance is maximized.

\[mev_i^j(P, s)=\max_{a_1 ... a_n \in A_j}\{ev_i(P, s, a_1... a_n)\}\]

Note that we introduced a new superscript, which binds the action set available to the player (in which domains the player can act). $mev^j_i$ therefore can be interpreted as the MEV that can be extracted from a balance change in domain $i$, given actions in domain $j$.  The single-domain case is therefore the case when $i=j$, and the player extracting value acts in the same domain in which the value is also extracted:

\[mev_i^i(P, s)=\max_{a_1 ... a_n \in A_i}\{ev_i(P, s, a_1... a_n)\}\]

It is however also useful to reason about cases where $i \neq j$. In our model, state is not domain-specific, and domains can read and write to each others' states if allowed to in the action space. One example of this would be any Layer 2 (L2) protocol, which reads ETH state to check for deposits and fraud proofs. If we thus consider the case where $i$ is a Layer 2 protocol and $j$ is Ethereum, we can ask questions such as "what is the maximum one can extract inside the domain of the L2 through manipulating L1 state only", which may be useful for quantifying and isolating the economic effect of cross-domain state interactions.

Note also that there is no strict or formal distinction in the state space of our two domains. Our definition models the state of the world as a monolithic entity, considering a moment in time in which a player $P$ is able to act and affect one or both domains. We model the separation between domains and their trust guarantees by restricting the action space. For example, applying mempool transactions on ETH (a set of actions) will naturally affect only ETH-native state, scoping the influence of a particular player to a particular domain. We intentionally allow for fuzzier distinctions in actions that affect multiple domains simultaneously, to allow for modeling cross-chain communication protocols, bridges, and other interactions between domains as their own actions that act simultaneously on multiple domains.

\hypertarget{cross-domain-maximal-extractable-value}{%
\subsubsection{2.4 Cross-Domain Maximal Extractable
Value}\label{cross-domain-maximal-extractable-value}}

\hypertarget{two-domain-mev}{%
\paragraph{2.4.1 Two-Domain MEV}\label{two-domain-mev}}

Let's consider now domains \(i\) and \(j\), and for simplicity assume
there is a single entity controlling the sequencing of both domains for
the moment we're considering (generalizing MEV to include a notion of time beyond the instantaneous is left to future work).

What that entity is informally looking for is to maximize its balance across both domains. Otherwise stated, given its account
balance on each domain \(b_i\) and \(b_j\), what are the sequences of
actions in \(A_i\) and \(A_j\) that will give the entity the highest sum
\((b_i+b_j)\)?

More formally, we define the cross-domain maximal extractable value of
domains \(i,j\) as follows:

\[mev_{i,j}^{i,j}(P, s)=\max_{a_1 ... a_n \in A_i \cup A_j}\{ev_i(P, s, a_1... a_n) + p_{j\rightarrow i} (ev_j(P, s, a_1... a_n))\}\]

The key insight here is that what happens after a state change in domain
\emph{j} is that the set of \emph{reachable states} in domain \emph{i}
might change. A simple example is how a user's deposit in an L1 might
unlock funds in an L2, allowing now a new state in the L2 where the
users's balance is larger. We model this by allowing these domains to share a single state,
and to access each other's state. While most mempool actions will act only on the state of that domain,
some actions, such as bridging funds, may cause the state of one domain to affect the state of another.

Note that we introduce a pricing function \(p_{j\rightarrow i}\), which
helps us translate native balances across different domains. We compute
the balance in domain \emph{j} in units of the native asset of domain
\emph{i} by simply multiplying it by this pricing function.  This is required as to meaningfully add the balances in two domains, some notion of their equivalent conversion prices must exist. 

In our previous examples in Figures~\ref{fig:2domain},\ref{fig:3domain}, we would assume a 1:1 pricing of MATIC and WMATIC, with an action space on Ethereum including performing a Uniswap transaction, and an action space on Polygon/BSC including bridging ETH into the equivalent wrapped assets as well as performing Uniswap trades on each domain. In Figure~\ref{fig:2domain}, the balance on Ethereum would decrease by 238172.18 MATIC, and the balance on Polygon would increase by 288033.14 WMATIC. Plugging into the above formula, we obtain an Extractable Value of $-238172.18 + (1.0) 288033.14 = 49860$, as in the example. Interestingly enough, we note that even if we value WMATIC at a significant discount to MATIC, say at 90\%, due to bridging costs, an MEV opportunity still exists. In this case, the total profit will be $-238172.18 + (0.9) * 288033.14 = 21057$ MATIC, still a tidy profit of approximately \$35,600.

We further assume that \(p_{i\rightarrow j}=\frac{1}{p_{j\rightarrow i}}\). This means that exchange rate of inverse exchanges are multiplicative inverses, and allows us to simplify the above definition by removing a corresponding term inside our maximization where profit is denominated in asset $j$. 
While this property is guaranteed if assets share a single global orderbook, it may not hold for all decentralized assets.  More complex pricing curves may be considered as part of future work, including pricing models that are dependent on quantity exchanged, modeling supply/demand under complex liquidity conditions with higher accuracy.  We also assume $p_{i\rightarrow i} = 1$, that is that there is no cost to convert an asset into itself.

\hypertarget{Generalizing to N-Domains}{%
\paragraph{2.4.2 generalizing}\label{generalizing}}

We can generalize our 2-domain definition to a n-domain definition, looking at a player $P$ that has access to an action space representing abilities across multiple domains $A = A_1 \cup A_2 \cup ... A_n$, and that wants to maximize balances across domains $B = B_1 \cup B_2 \cup ... B_n$ with respective pricing functions of each asset priced in domain $B_1$ as $p_{B_1\rightarrow B_1}, ..., p_{B_n\rightarrow B_1}$\footnote{$B_1$ is chosen as the canonical base asset without loss of generality, due to the inverse property of exchange. Generalizing this definition to pricing functions without the inverse property would require summing across all possible base assets, or perhaps inventing a synthetic base asset which represents the relative desirability of each domain's asset.}:

\[mev^A_{B}(P,s)=\max_{a_1 ... a_n \in A}\{\sum_{b\in B} p_{b\rightarrow B_1}(ev_b(P, s, a_1... a_n))\}\]

We can see that the MEV is the maximum of the sum of final balances across all considered domains into a single base asset (canonically the first domain considered), when some mix of actions across all those domains are executed together.

We can also reason about the total MEV available to extract in the world, by allowing $A$ to represent the joint action-space of all domain sequencers (as well as any protocol bridges and the expected action-space of cross-domain communication infrastructure), and by allowing $B$ to represent all assets on all domains.

Having defined cross-domain MEV with full generality, we will now reason about the logical implications of this definition on blockchain protocols, and prescribe important areas for future study that are likely to be impactful as a multi-chain future proliferates.

\hypertarget{sequencer-collusion}{%
\subsection{3. Sequencer collusion}\label{sequencer-collusion}}

In our 2-domain case, we assumed for simplicity that a single entity controls
the ordering on both domains. In reality, there will likely be different
sequencers for each domain.

MEV extraction has historically been thought of as a self-contained
process on a single domain, with a single actor (traditionally the miner) earning an atomic profit that serves as an implicit transaction fee. In a multi-chain future, extracting the maximum possible value from multiple domains
will likely require collaboration, or collusion, of each domain's
sequencers if action across multiple domains is required to maximize profit.

To analyze this, we introduce a term $\alpha$, which represents the cost in the base asset for a set of multiple sequencers across multiple domains to collude. The incentive for two sequencers to collude can be split into three
categories:

\begin{itemize}
\item
  if \(mev_{i,j}^{i,j}(s) > mev_{i}^{i}(s)+mev_{j}^{j}(s)+\alpha\), then
  the sequencers make more value through collusion, as the benefit of collusion over acting independently outweighs the cost $\alpha$.
\item
  if \(mev_{i,j}^{i,j}(s) = mev_{i}^{i}(s)+mev_{j}^{j}(s)+\alpha\), then
  there is no difference between colluding and not doing so.
\item
  if \(mev_{i,j}^{i,j}(s)< mev_{i}^{i}(s)+mev_{j}^{j}(s)+\alpha\), then
  the sequencers do not make more value by colluding.
\end{itemize}

\textbf{Example}. Suppose there exist two automated market makers with some (potentially different) on-chain pricing function: Uniswap and Toroswap.
They are both markets between ETH and DAI. Uniswap is on domain \(i\),
and Toroswap is on domain \(j\).

Suppose they have the same amount of liquidity and are both indicating a
price of 20 DAI/ETH. Further suppose there is no other activity on each of these
domains.

Let a single large buy-ETH transaction push the Uniswap market to 30 DAI/ETH.

Since Toroswap is still at 20 DAI/ETH, there exists an arbitrage opportunity
between Uniswap and Toroswap, that will result in each pool indicating a
price of 25. Further assume the profit from this arbitrage opportunity
is 1 eth and that the cost of collusion \(\alpha\) is 0.

In this example, we have $A_i$ consisting of making a Uniswap trade, and $A_j$ consisting of making a Toroswap trade. We will assume the player $P$ has enough balance in each state to perform its choice of trade (aka that $P$'s balance exceeds the arbitrage opportunity).  We now have that:

\(mev_{i}^{i}(s)=\max_{a_1 .. a_n \in A_i}\{ev_i(s, a_1 ... a_n)\}=0\)

\(mev_{j}^{j}(s)=\max_{a_1 .. a_n \in A_j}\{ev_j(s, a_1 ... a_n)\}=0\)

\(mev_{i,j}^{i,j}(s)=\max_{a_1 ... a_n \in A_j \cup A_i}\{ev_i(s, a_1 ... a_n) + p_{j \rightarrow i}ev_j(s, a_1 ... a_n)\}= 1 \ eth\)

In the last case, performing an arbitrage trade as described in the 2-AMM case in~\cite{babel2021clockwork} results in a guaranteed profit, increasing a user's ETH balance if it can perform trades atomically across both AMMs.

So we've found an example of a case where the first inequality laid out above, \(mev_{i,j}^{i,j}(s) > mev_{i}^{i}(s)+mev_{j}^{j}(s)\), holds. In Appendix B, we consider an illustrated example with 4 AMMs, two in \(i\)
and two in \(j\).

We expect that, given the deployment of AMMs and other MEV-laden technologies across multiple domains, the benefit of extracting MEV across multiple domains will often outweigh the cost of collusion $\alpha$.

\hypertarget{trade-mechanics-market-structure}{%
\subsubsection{3.1 Trade Mechanics \& Market
Structure}\label{trade-mechanics-market-structure}}

So far, we've ignored the mechanics of actually seizing such an
opportunity. Since it exists across domains and given it is finite, the
competition for such opportunities will be fierce and it is likely no bridge will be fast enough to execute a complete arbitrage transaction as exemplified in Figure~\ref{fig:2domain}.

One observation is that a player that already has assets across both domains does not need to bridge funds to capture this MEV profit, reducing the time, complexity, and trust required in the transaction. This means that cross-domain opportunities may be seized in two simultaneous
transactions, with inventory management across many domains being internal to a player's strategy to optimize their MEV rewards.

Such behavior is similar to the practice of inventory management that
market makers and bridges in traditional finance do, which primarily consists of keeping assets scattered across multiple heterogeneous-trust domains (typically centralized exchanges), managing risks associated with these domains, and determining relative pricing. Some key differences include the ability to coordinate with actors in the system other than themselves, such as through a DAO or a system like Flashbots. However, given these conclusions, it is likely that traditional financial actors may have a knowledge-based advantage in cross-domain MEV risk management, which may induce centralization vectors that come from such actors being able to run more profitable validators.

Despite being grim, this fact is important as it reveals a key property that
cross-domain interactions are subject to: the loss of composability.
There is no more atomic execution. This introduces additional execution
risk, as well as requires higher capital requirements, further raising the barriers-to-entry required to extract MEV. We expect bridges to play an extremely important role in such an MEV ecosystem, as the cheaper, more ubiquitous, and faster bridges become, the more competitive these arbitrage transactions naturally become by decreasing the inequality of the action space across players as a function of their capital.

\hypertarget{cost-of-collusion}{%
\subsubsection{3.2 Where does the cost of collusion come from?}\label{cost-of-collusion}}

Assuming the sequencers of \(i\) and \(j\) are distinct, they need a way
to communicate and trust each other in order to apply the necessary
orderings to their respective domains to maximize profit and split
rewards between themselves.

From past results in~\cite{zamyatin2021sok}
\footnote{described in \url{https://ls.mirror.xyz/5FM0KUurAEkN7yDXLAI8i9kCha\_yMlzguVqiHEGMjPU}},
we know that cross-chain communication is impossible without either a
trusted third party or a synchrony assumption other than asynchrony.  Since ordering is time-sensitive, it is likely a synchrony assumption
cannot be made and so a trusted third party is required to access these MEV opportunities. The presence of a trusted third-party can be thought about as an additional cost to facilitate cross-domain collusion.

However, if the sequencers of \(i\) and \(j\) are controlled by a single
entity, then no trust is needed between them and this cost is now close
to 0. This could create unwanted behaviours where entities will seek
voting power across domains in order to bring their cost of cross-domain
MEV extraction (\(\alpha\)) down. This also means that colluding sequencers will naturally have an advantage over single-chain sequencers in the MEV market, as they will have access to a wider action space. It is worth noting that in practice, mechanisms like Flashbots or an SGX-based DAO may lower the cost of collusion even for single-chain sequencers.  While these systems provide some promise to allow non-colluding sequencers to remain profitable, these systems require additional trust guarantees. If the profit from colluding is substantial, it is likely such collusion mechanisms will be used in practice, becoming de-facto defaults.

Another cost which we have not considered is the indirect cost of
collusion. Even in the cases where \(mev_{i,j}^{i,j}(s)\) is far greater than
\(mev_{i}^{i}(s)+mev_{j}^{j}(s)+\alpha\), there may exist social norms
against colluding.

If sequencers were to infringe on these norms, one could imagine they could be punished by the
community or that the domain's token price could go down as trust in the
domain's fairness diminishes, which could impact the sequencer's
holdings. It may be possible to estimate this implicit cost by observing cases where entities have access to cross-domain MEV, and observing at which threshold of market growth they begin extraction activities.

\hypertarget{negative-externalities}{%
\subsection{4 Negative externalities}\label{negative-externalities}}

Cross-domain extractable value might surface new negative externalities
the community should be aware of. While they warrant further study, we
introduce some of them here:

\textbf{Cross-domain sequencer centralization}

Cross-domain extractable value may create an incentive for sequencers
(i.e.~validators in most domains) to amass votes across the networks
with the most extractable value.

This is especially relevant when realizing there already exist large validators and staking providers running infrastructure across many networks\linkto{https://www.stakingrewards.com/providers/}. It is unlikely that those who are for-profit entities will forego access to MEV revenue long term if such revenue is substantial.

\textbf{Cross-domain time bandit attacks}

Time-bandit attacks were first introduced in~\cite{daian2020flash},
and consist in looking at cases where the miner
has a direct financial incentive to re-org the chain it is mining on.

In a cross-domain setting, there may now exist incentives to re-org
multiple domains, or to re-org weaker domains in order to execute
attacks resembling double-spends. This will be particularly relevant to
the security of bridges.

\textbf{Super-traders}

While the risks above are worrisome, historically the negative
externalities around extractable value have surfaced not from the
sequencers' (miners) behaviour but from the dynamics that the pursuit of
this value create in the markets.

One general worry is of the potential economies of scale, or moats, that
a trader could create across domains which would end up increasing the
barrier to entry for new entrants and enshrine existing players'
dominance.

\textbf{example.} Suppose a domain orders transactions on a first-in
first-out basis (FIFO). Such a domain effectively creates a latency race
between traders going after the same opportunity. As seen in traditional
markets, traders will likely invest in latency infrastructure in order
to stay competitive, possibly reducing the efficiency of the market~\cite{budish2015high}.

If several domains also have such ordering rules, traders
could engage in a latency race in each of them, or a latency race to propagate arbitrage across domains, making the geographical points in which these systems operate targets for latency arbitrage. It is likely the
infrastructure developed to optimize one domain, can be used across multiple domains.
This could create a `super' latency player, and certainly advantages entities which already have considerable expertise in building such systems in traditional finance. Such latency-sensitive systems may erode the security of systems that do not rely on latency, by advantaging latency-optimizing players in the cross-domain-MEV game. This area warrants substantial further study, as it is well-known that global network delays require relatively long block times in Nakamoto-style protocols to achieve security under asynchrony~\cite{pass2017analysis}, and it is possible cross-domain-MEV may erode the fairness of validator rewards in such protocols.

\textbf{Subjectivity of MEV}

One important consequence of this definition is the introduction of heterogeneous pricing models across different players in the blockchain ecosystem. Previously, MEV was an unambiguous quantity denominated in a common base asset, ETH. In a multi-chain future, the relative price differences between actors is not only relevant in calculating MEV, it can in fact \emph{create} MEV.  For example, a previous validator may leave a system in what is in their pricing model a 0-MEV state, but the next validator, who disagrees with this pricing model, may see opportunities to rebalance MEV to increase assets it subjectively values more. This provides yet another intution for why MEV is fundamental to global, permissionless systems even if they do not suffer from ordering manipulation of the kind described in~\cite{daian2020flash}.

\hypertarget{open-questions}{%
\subsection{5 Open questions}\label{open-questions}}

We end with a set of open questions. These are questions
we're thinking about and looking to collaborate on.

\textbf{How do we best define the action space?}

Astute readers will note that we have punted many hard problems, including defining the boundaries between contracts on various domains, modeling the trust assumptions of oracles and bridges, and modeling probabilistic interactions, e.g. between centralized and decentralized exchanges. While some of this ambiguity is intentional, we believe it is also manageable.~\cite{babel2021clockwork} provides one example for how the model in this paper can be instantiated in any transaction-based blockchain setting, including Ethereum, leaderless L1 protocols, Layer 2 protocols, and more. One natural application of this work is to extend the models therein to include other Ethereum-style domains, with action spaces defined similarly. This would prove sufficient for reasoning about much of the MEV we have explored in this work. We leave more complex modeling to future work, and would like this work to open discussion about what the most useful models would be in practice. We are optimistic that an open-source, mathematically formal, and executable set of models as proposed in~\cite{babel2021clockwork} are a feasible path for the community to rigorously tackle MEV going forward.

\textbf{Aside from cross-domain arbitrage, what are other forms of
cross-domain MEV?}

In the near future, there will be cross-domain smart contract calls
enabling applications such as cross-domain lending and voting. Is there
any new cross-domain extractable value that will be created from these
applications? What about the extractable value created from bridges, oracles, governance, and other cross-domain systems?

\textbf{What does a protocol for sequencer collusion look like and what
are its desirable properties?}

If cross-domain MEV extraction is inevitable, what does a good mechanism
look like for it to be extracted? More concretely, how can two
sequencers that do not trust each other share information about the
state, and share profits from their collaboration, across domains with
different finality and consensus mechanisms?

\textbf{How can we identify and quantify cross-chain MEV extraction
taking place?}

While \linkto{https://westerngate.xyz}{Westerngate.xyz} does a great job
at identifying potential cross-chain arbs, identifying historical
extraction in practice seems a lot more difficult. How can we do so? If
it is not possible, are we headed into a world where this activity is a
lot more opaque than it has been so far?

\textbf{What can we learn from existing distributed and parallel
programming literature?}

The concept of a domain extends beyond the world of cryptocurrency, and
finds an analogue in classical parallel/distributed programming. The
differences arise in the fact the systems we study need to consider
adversarial behavior. Are there existing results from these subjects'
literature that we can re-use or inspire ourselves of? In Appendix A we
dive into this in further detail.

\begin{center}\rule{0.5\linewidth}{0.5pt}\end{center}

\section*{Acknowledgements}

Thanks to \linkto{https://flashbots.net}{Flashbots} for supporting and funding this research. 

Thanks to Vitalik Buterin, Kushal Babel, Patrick McCorry and Georgios
Konstantopoulos for commenting on versions of this work, and to the
Flashbots crew for many stimulating conversations that lead to it.

\bibliographystyle{plain} 
\bibliography{refs} 

\hypertarget{appendix}{%
\subsection{Appendix:}\label{appendix}}

\hypertarget{a.-comparison-to-classical-parallel-programming}{%
\subsubsection{A. Comparison to classical parallel
programming}\label{a.-comparison-to-classical-parallel-programming}}

The concept of a domain extends beyond the world of cryptocurrency, and
finds an analogue in classical parallel/distributed programming.

In this Appendix, we briefly lay out the similarities and differences
between traditional distributed systems concepts, and their analogues in
censorship-resistant distributed systems. By doing so, our hope is that
this will open up the questions here to a wider audience, who will be
able to draw parallels with existing research from their field.

We will start by looking at the program execution model in Linux and map
existing abstractions to domains.

In Linux, the default self-contained unit for executing transactions in
a synchronous manner is a \emph{thread}. Threads are created by a
\emph{process}, which manages the stored state and synchronization
amongst different threads. The \emph{kernel} manages the set of
processes and handles scheduling process execution and mappings of
virtualized state and bytecode to hardware.

Domains are most like processes in that they manage the shared state
amongst a number of contracts. Individual contracts can be thought of as
threads and the execution model of whether threads are run sequentially
or in parallel depends on the host chain.

\textbf{Example 1.} In Solana, one can provide a dependency graph
between contracts and contracts in different strongly connected
components of this dependency graph are executed in parallel.

\textbf{Example 2.} On the other hand, in the current Ethereum Virtual
Machine (EVM), contracts are always executed sequentially with the order
determined by the sequencer.

In classical parallel/distributed programming, one often does not make a
distinction between the process (domain) and the kernel that executes
the process (sequencer). However, in censorship-resistant decentralized
systems we are forced to separate these two to ensure that we minimize
excess profit taken by the execution layer.

Part of the aim of this paper is to provide some insight into the case where there are multiple sequencers (e.g.~multiple
kernels) that are interacting.

Multiple domains (processes) communicate via a communication channel that
sends messages whose validity can be quickly verified via a
cryptographic hash or commitment. Cross-domain communication is
analogous to
\linkto{https://en.wikipedia.org/wiki/Futures\_and\_promises}{futures and
promises} from traditional distributed programming, where one process or
kernel submits a remote deferred computation and waits for a response.
Again, in order to guarantee censorship resistance, we have to separate
the actual message sent from the entity executing the relay of the
message. One of the main differences between traditional deferred
computation and what is found in cryptocurrencies is the excess overhead
for verifying computation on both sides. We will provide a brief
overview, but please read
\linkto{https://stonecoldpat.github.io/images/validatingbridges.pdf}{this
review article} for more details.

Currently, the best solutions in cryptocurrency for cross-domain
communications are \emph{bridges}. Bridges involve three principal
agents: 

\begin{itemize}
  \item Sequencer of the source chain
  \item Sequencer of the destination chain
  \item Relayer of messages from the source and destination chain
\end{itemize}

The sequencer of the source chain produces the outbound transaction and
verifies it in the bridge contract. The bridge contract effectively acts
as the store of deferred computation (almost like a
\linkto{https://en.wikipedia.org/wiki/Thunk}{thunk}). The relayer runs
blockchain clients for both the source and destination chain and upon
finding a valid ``forward'' event from the source chain, submits a
transaction on the destination chain. The sequencers of the destination
chain validate the forward event in the receive side bridge contract.
Upon being validated, applications can interact with the relayed
message. In order to guarantee that a relayer cannot send an invalid
message or censor a valid message, cryptographic and economic techniques
are utilized.

One can view the communication complexity as similar to the synchronization cost that one naturally has to deal with in parallel programming. For instance, you might have n threads, but your algorithm needs $\sqrt{n}$ queues/spinlocks to handle communication, so you only end up with a $\sqrt{n}$ improvement. However, unlike the parallel programming scenario, there is now an adversarial component to how your distribute your computation and communication. One needs to adjust their threat model to include that adversaries will grief the user by forcing them to pay more than necessary or will increase latency dramatically.

\hypertarget{b.-4-amms-example}{%
\subsubsection{B. 4-AMMs case
example}\label{b.-4-amms-example}}

While our 2-AMM example in section 3 above presents a simple case where the inequality holds. It does not consider any activity within each domain that could conflict with the realization of a cross-domain opportunity.

In this example, we show how for price inefficiencies for AMMs
within and across domains, the inequality still holds.

\textbf{example}. Suppose there exist four market makers with some (potentially different) on-chain pricing function: Uniswap, Sushiswap,
Toroswap and Unagiswap. They are all markets between ETH and DAI. The
first two are on domain \(i\), and the second two are on domain \(j\).

Suppose they all have the same amount of liquidity and are all
indicating a price of 20 DAI/ETH. Further suppose there is no activity on each of these domains aside from DEX trades and arbitrage trades.

Let a single large buy-ETH transaction \(tx_1^i\) push the Uniswap
market to 30.

Since Sushiswap is still at 20 DAI/ETH, Uniswap can now be rebalanced through
arbitrage to 25 via transactions \(tx_2^i\) and \(tx_3^i\), so that
Uniswap and Sushiswap pools will indicate a price of 25. Further assume the profi t from this arbitrage opportunity is 1 eth and that the cost of collusion \(\alpha\) is 0. 

However, Toroswap and Unagi are still at 20, and so they can be further
rebalanced through arbitrage between them and (Uni,Sushi) via
transactions \(tx_4^i,tx_5^i,tx_1^j,tx_2^j\), such that now all markets
are at 22.5, netting an additional profit of 0.3 eth for each
rebalancing.

In this example, we have $A_i$ consisting of making a Uniswap trade, and $A_j$ consisting of making a Toroswap trade. We will assume the player $P$ has enough balance in each state to perform its choice of trade (aka that $P$'s balance exceeds the arbitrage opportunity).  We now have that:

\begin{itemize}
\tightlist
\item
  \(tx_1^i\): is an ETH-buy transaction on Uniswap in \(i\)
\item
  \(tx_2^i\): is a ETH-buy transaction on SushiSwap in \(i\)
\item
  \(tx_3^i\): is a ETH-sell transaction on Uniswap in \(i\)
\item
  \(tx_4^i\): an ETH-sell transaction on Uniswap in \(i\)
\item
  \(tx_5^i\): an ETH-sell transaction on Sushiswap in \(i\)
\item
  \(tx_1^j\): an ETH-buy transaction on UnagiSwap in \(j\)
\item
  \(tx_2^j\): an ETH-buy transaction on Toroswap in \(j\)
\end{itemize}

\begin{figure}[H]
    \centering
    \includegraphics{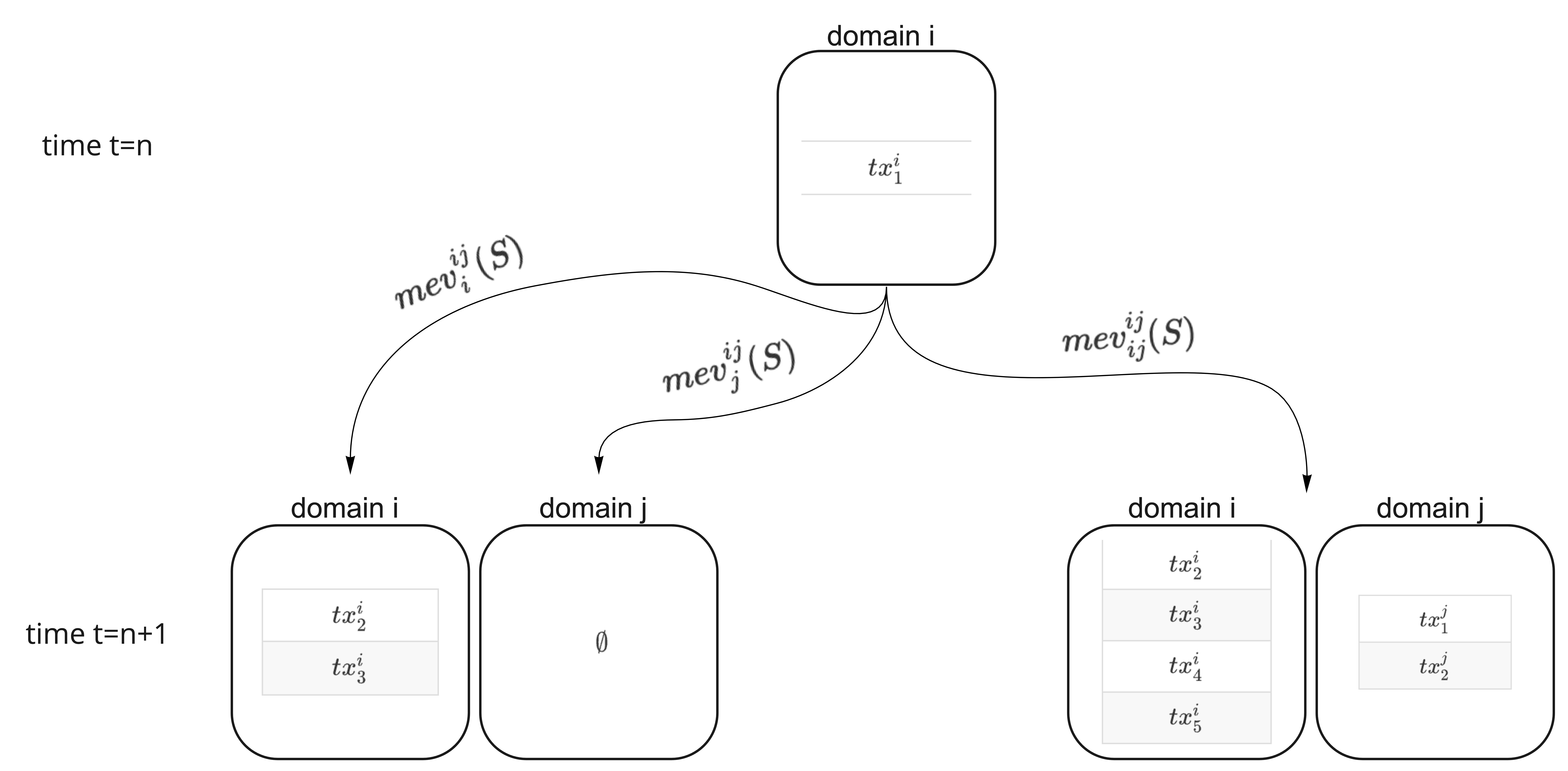}
    \caption{Example of multi-AMM multi-domain MEV under collusion. The left side shows each domain optimized for MEV individually, while the right side shows transaction optimization under sequencer collusion.}
    \label{fig:multi}
\end{figure}

\(mev_{i}^{i}(s) = 1 \ eth\)

\(mev_{j}^{j}(s) = 0 \ eth\)

\(mev_{i,j}^{i,j}(s) = 1+0.3+0.3=1.6 \ eth\)

so we've found a case where:

\(mev_{i,j}^{i,j}(s) > mev_{i}^{i}(s)+mev_{j}^{j}(s)\)

\hypertarget{caveats}{%
\paragraph{Caveats}\label{caveats}}

In addition, and again for clarity, our example considers next-block
arbitrage rather than backrunning, and ignores other transactions in
each domain blocks that may be conflicting with each other.

While this example is simplified, one can see how this holds in more
complex settings for cross-domain arbitrage. Since arbitrage is the bulk
of MEV extraction today, one can see how such a scenario may happen
regularly.
\end{document}